\RequirePackage{fix-cm}

\documentclass[twocolumn,epjc3]{svjour3}  
\usepackage{lineno}

\smartqed  

\RequirePackage{graphicx}
\RequirePackage{amsmath}
\RequirePackage{xspace}
\RequirePackage{booktabs}

\newcommand*{\HWW}{\ensuremath{\text{H}\!\to\!\text{WW}^*}\xspace}
\newcommand*{\HZZ}{\ensuremath{\text{H}\!\to\!\text{ZZ}^*}\xspace}
\newcommand*{\WW}{\ensuremath{\text{WW*}}\xspace}
\newcommand*{\ZZ}{\ensuremath{\text{ZZ*}}\xspace}

\newcommand*{\GeV}{\xspace\ensuremath{\text{Ge\kern-0.1em V}}\xspace}
\newcommand*{\MeV}{\xspace\ensuremath{\text{Me\kern-0.1em V}}\xspace}

\newcommand*{\Z}{\ensuremath{\text{Z}}\xspace} 
\newcommand*{\Zg}{\ensuremath{\text{Z/}\gamma}\xspace}
\newcommand*{\invab}{\ensuremath{\text{ab}^{-1}}\xspace}
\newcommand*{\invfb}{\ensuremath{\text{fb}^{-1}}\xspace}

\newcommand*{\WHIZ}{{\tt WHIZARD3}\xspace}

\newcommand*{\PYTHIAs}{{\tt PYTHIA6}\xspace}
\newcommand*{\PYTHIAe}{{\tt PYTHIA8}\xspace}
\newcommand*{\DELPHES}{{\tt DELPHES}\xspace}
\newcommand*{\KEYfHEP}{{\tt Key4HEP}\xspace}
\newcommand*{\FASTJET}{{\tt FastJet}\xspace}

\newcommand*{\sqrts}{\ensuremath{\sqrt{s}}\xspace}
\newcommand*{\sqrtsZH}{\ensuremath{\sqrts = 240~\GeV}\xspace}

\usepackage{microtype}
\journalname{Eur. Phys. J. C}

\begin{document}

\title{Simultaneous measurement of the Higgs boson decay rates to \WW and \ZZ in fully hadronic final states at FCC-ee}

\author{
Jan Eysermans\thanksref{addr1,addr2}
\and
Mila Pandurovic\thanksref{addr3}
\and
Christoph Paus\thanksref{addr1}
\and
Michele Selvaggi\thanksref{addr2}
\and
Aman Desai\thanksref{addr4}
}


\institute{Particle Physics Collaboration, Massachusetts Institute of Technology,
77 Massachusetts Ave, Cambridge, MA 02139, USA \label{addr1}
\and
CERN, 1211 Route de Meyrin, 1211 Geneva, Switzerland \label{addr2}
\and
Vinca Institute of Nuclear Sciences,
Mike Petrovica Alasa 12--14, 11090 Belgrade, Serbia \label{addr3}
\and
Adelaide University, Adelaide, Australia \label{addr4}
}



\date{Received: date / Accepted: date}

\maketitle

\begin{abstract}

The precise determination of the Higgs boson couplings to electroweak gauge bosons is a key objective of the FCC-ee physics program. In particular, measurements of the Higgs boson couplings to W and Z bosons play a central role in the model-independent determination of its total decay width. This paper presents a study of Higgs boson decays to \WW and \ZZ in the Higgsstrahlung production process. The analysis focuses on events in which the associated \Z boson and the vector bosons from the Higgs decay all decay hadronically, resulting in six-jet final states. The large \HWW branching fraction leads to high signal statistics in the fully hadronic channel, while for \HZZ the much smaller Higgs branching fraction is partly compensated by the large hadronic branching fractions of the Z bosons. Due to the substantial kinematic overlap between the \WW and \ZZ decay modes, their signal strengths are extracted simultaneously. Jet-clustering and dedicated jet-pairing techniques are employed to optimize the reconstruction of the hadronically decaying gauge bosons and enhance the separation of the two signal components. Assuming a center-of-mass energy of 240~GeV and an integrated luminosity of 10.8~\invab, expected relative uncertainties of 1.48\% and 8.20\% are obtained on the production cross section times branching fraction, $\sigma_{\mathrm{ZH}}\times\mathrm{BR}$, for the \HWW and \HZZ decay modes, respectively. These estimates include background-normalization uncertainties; detector-modeling and other experimental systematic uncertainties are not evaluated but are expected to have a subdominant impact.

\end{abstract}

\section{Introduction} \label{intro}

The discovery of the Higgs boson at the Large Hadron Collider (LHC)~\cite{HIGG-2012-27,CMS-HIG-12-028} completed the particle content of the Standard Model (SM), yet several fundamental questions, including the nature of dark matter and the origin of the matter--antimatter asymmetry, remain unanswered. Searches for new particles have so far revealed no significant deviations from the SM, motivating complementary searches through precision measurements. In many scenarios of physics beyond the SM (BSM), including extended scalar sectors~\cite{PhysRevD.8.1226} and composite Higgs models~\cite{KAPLAN1984183}, deviations of the Higgs boson couplings from their SM predictions can arise at the percent level or below. The High-Luminosity LHC (HL-LHC) is expected to substantially improve the precision of Higgs measurements~\cite{atlas2025highlightshllhcphysicsprojections}, reaching the percent level for several couplings. At an electron--positron collider, the combination of inclusive and exclusive Higgs measurements can further improve the precision of selected Higgs couplings to the per-mille level in a model-independent global fit. This is enabled by the clean collision environment and well-defined initial state, which allow for precise reconstruction of the Higgs decay products.

The Future Circular Collider (FCC) program~\cite{FCC:2025lpp} is planned in two stages, beginning with the electron--positron collider (FCC-ee) and followed by the high-energy hadron collider (FCC-hh). FCC-ee is planned to operate at several center-of-mass energies (\sqrts) to perform precision measurements at the Z pole, the WW production threshold, the Higgs production stage around 240~GeV, and the top-quark pair production run up to about 365~GeV. At \sqrtsZH, Higgs production is dominated by the Higgsstrahlung process, $\rm e^+e^- \to ZH$, yielding a sample of $\mathcal{O}(10^6)$ Higgs bosons for an integrated luminosity of 10.8~\invab accumulated across the four interaction points. At $\sqrt{s}=365$~GeV, Higgsstrahlung remains the dominant production mode, while the WW-fusion process, $\rm e^+e^- \to H\nu\bar{\nu}$, reaches approximately one quarter of the Higgsstrahlung yield.

A central objective of the FCC-ee Higgs program is the precise determination of the Higgs couplings to SM particles. The couplings to the electroweak gauge bosons, $\rm g_Z$ and $\rm g_W$, are particularly important observables, providing sensitive probes of possible deviations from the SM and playing a central role in the model-independent determination of the absolute Higgs couplings and total Higgs boson width $\rm \Gamma_H$~\cite{FCC:2025lpp}. The inclusive Higgsstrahlung cross section, $\rm \sigma_{ZH}$, can be measured with the recoil-mass technique independently of the Higgs decay mode and directly constrains the absolute $\rm g_Z$ coupling through $\rm \sigma_{ZH}\propto g_Z^2$. At FCC-ee, $\rm \sigma_{ZH}$ can be determined with per-mille-level precision~\cite{li2025modelindependentzhproductioncross}. Measurements of exclusive Higgs final states provide complementary sensitivity to the Higgs couplings and total width. For a generic decay $\rm H\to X$, the corresponding rates in Higgsstrahlung and WW-fusion production scale as
\begin{equation}
\rm \sigma_{ZH}\times BR (H\to X) \propto \frac{g_{Z}^{2}g_{X}^{2}}{\Gamma_H},
\end{equation}
and
\begin{equation}
\rm\sigma_{H\nu\nu}\times BR (H\to X) \propto \frac{g_{W}^{2}g_{X}^{2}}{\Gamma_H}, \label{eq:fusion}
\end{equation}
respectively. In particular, the $\rm \sigma_{ZH}\times BR (\HZZ)$ rate scales as $\rm g_Z^{4}/\Gamma_H$ and, when combined with the model-independent determination of $\rm g_Z$ from the inclusive Higgsstrahlung cross section, provides an absolute constraint on $\rm\Gamma_H$. The \HWW decay provides direct sensitivity to $\rm g_W$ through the Higgs decay vertex. When combined with the enhanced WW-fusion contribution at $\sqrt{s}=365$~GeV, which probes $\rm g_W$ through the production vertex, the \HWW channel provides important complementary sensitivity to $\rm \Gamma_H$ and substantially improves the expected precision compared with the 240~GeV data alone. Within the broader FCC-ee Higgs program, the combination of inclusive and exclusive Higgs production and decay channels at both $\sqrts=240$ and 365~GeV allows $\rm \Gamma_H$ to be determined with an expected precision of about 0.78\%~\cite{FCC:2025lpp}.

In this work, we investigate the \HWW and \HZZ decay modes in associated Higgs production at \sqrtsZH, focusing on the fully hadronic final state in which the associated Z boson and the vector bosons from the Higgs decay into quarks. In the SM, the corresponding Higgs-boson branching fractions are approximately $\rm BR(\HWW)\approx21.5\%$ and $\rm BR(\HZZ)\approx2.6\%$, such that the \HZZ decay mode is about eight times less abundant than \HWW. The large hadronic branching fractions, $\rm BR(Z\to q\bar{q})\approx0.7$ and $\rm BR(W\to q\bar{q}')\approx0.67$, provide substantially larger signal yields for both \HWW and \HZZ than the corresponding final states involving leptonic vector-boson decays. At the same time, the resulting multijet topology presents significant experimental challenges, including large diboson and fermion-pair backgrounds, combinatorial ambiguities in the assignment of jets to their parent bosons, and the separation of \HWW and \HZZ events.

Complementary leptonic, semileptonic, and invisible final states have also been studied at FCC-ee, providing additional sensitivity to the \HWW and \HZZ decay rates~\cite{Selvaggi:2025kmd}. Their combination with the fully hadronic channels studied here is discussed later together with the presentation of the results.

Similar studies of the fully hadronic $\rm ZH(\HWW)$ final state have been performed for other proposed Higgs factories. At the ILC, a full-simulation study at $\sqrt{s}=250$~GeV with an integrated luminosity of 500~\invfb obtained an expected relative statistical uncertainty of 4.1\% on $\rm \sigma_{ZH}\times BR(H\to WW^*)$~\cite{Pandurovic:2019ILC}. A corresponding CEPC study of the fully hadronic final state at $\sqrt{s}=250$~GeV found an expected precision of 1.9\% for 5.0~\invab~\cite{Pandurovic:2020CEPC}, with a similar sensitivity to the present result when accounting for the difference in integrated luminosity.

This paper is organized as follows. Section~\ref{strategy} outlines the overall analysis strategy. The simulated event samples and detector response are described in Sections~\ref{evtgen} and~\ref{detresp}, respectively, followed by the event preselection in Section~\ref{preselection}. The separation of the \HWW and \HZZ signal processes is discussed in Section~\ref{hzz_hww_sep}, while the category-specific signal optimizations are presented in Section~\ref{optimize}. Finally, Section~\ref{results} presents the statistical interpretation and results, with the conclusions summarized in Section~\ref{conclusion}.

\section{Analysis strategy}\label{strategy}

This paper aims to extract the \HWW and \HZZ signal strengths in fully hadronic final states. Because of the substantial kinematic overlap between the hadronic \WW and \ZZ decay modes, their signal strengths are determined simultaneously. A key challenge is the separation of the \HWW and \HZZ contributions, whose fully hadronic final states exhibit very similar event topologies. The finite dijet-mass resolution further leads to a significant overlap between reconstructed W- and Z-boson candidates. This is particularly important for the \HZZ measurement, since its branching fraction is approximately eight times smaller than that of \HWW, making it more susceptible to contamination from misidentified \HWW events in the overlapping invariant-mass regions. An efficient separation of the two decay modes is therefore essential for the simultaneous determination of their signal strengths.

The analysis starts with a common preselection, described in Section~\ref{preselection}, which suppresses the dominant backgrounds while retaining both signal processes. The selected events are subsequently separated into mutually exclusive \HWW-like and \HZZ-like categories using a dedicated multivariate classifier, as described in Section~\ref{hzz_hww_sep}. The classifier combines reconstructed boson kinematics with jet-flavor information to discriminate between the two signal hypotheses.

Background rejection is then optimized independently within each category using dedicated multivariate discriminants constructed from the relevant event observables, as described in Section~\ref{optimize}. The \HWW and \HZZ signal strengths are finally extracted through a simultaneous fit to the discriminant distributions in both categories, accounting for signal cross-contamination, background contributions, and the associated uncertainties, as detailed in Section~\ref{results}.

\section{Event generation} \label{evtgen}

The Monte Carlo samples used in this analysis are generated at \sqrtsZH and normalized to an integrated luminosity of 10.8~\invab, as foreseen for the baseline FCC-ee running scenario with four interaction points. The collision conditions include a relative beam-energy spread of $0.185\%$ for each beam, corresponding to approximately 222~MeV, as well as a smearing of the interaction vertex according to the FCC-ee beam parameters documented in the Feasibility Study Report~\cite{FCC:2025lpp}.

Two event generators are used in this analysis depending on the process. The signal and fermion-pair samples are generated with \WHIZ~\cite{Kilian:2007gr}, while the diboson samples are generated with \PYTHIAe~\cite{Sjostrand:2014zea}. For the \WHIZ{} samples, the hard-scattering process is simulated at tree level and the generated events are interfaced to \PYTHIAs~\cite{Sjostrand:2006za} for parton showering and hadronization, as well as for Higgs-boson decays in the signal samples. Initial-state radiation and the beam-energy spread are included in \WHIZ{}, while initial-state radiation is disabled in the subsequent \PYTHIAs{} step. In samples generated with \PYTHIAe{}, the hard process, parton showering, and hadronization are all handled by the same generator. No fixed-order NLO QCD or electroweak corrections are applied to the hard-scattering processes.

The signal consists of associated Higgs-boson production, $\rm e^+e^- \to ZH$, with the analysis focusing on the \HWW and \HZZ decay modes. The associated Z boson is required to decay hadronically, with dedicated samples produced for the different quark flavors, including light-quark, charm, and bottom-quark final states. The Higgs boson is decayed into the \HWW and \HZZ final states, with both vector bosons subsequently decaying hadronically into quarks. Final states involving tau leptons, including their hadronic decays, are not considered, as their contribution to the selected signal-like events is expected to be negligible.

The dominant backgrounds arise from diboson and hadronic fermion-pair production. The $\rm e^+e^- \to W^+W^-$ sample, denoted as $\rm WW$ throughout this analysis, includes all W-boson decay modes. Neutral-current diboson production, denoted as $\rm ZZ$ throughout this analysis, is described by $\rm e^+e^- \to (\gamma^*/Z)(\gamma^*/Z)$ with inclusive decays of both neutral bosons. Consequently, this sample contains the $\rm ZZ$, $Z\gamma^*$, $\gamma^*Z$, and $\gamma^*\gamma^*$ contributions and their interference within the double-resonant approximation implemented in \PYTHIAe{}. Fermion-pair production, $\rm e^+e^- \to Z/\gamma^* \to q\bar{q}$, is restricted to hadronic two-fermion hard-process final states, with initial- and final-state photon radiation included.

Each simulated sample is normalized to the production cross section provided by the corresponding event generator. The samples considered in the analysis, together with their cross sections for the fully hadronic final states and expected event yields, are summarized in Table~\ref{tab:samples}.

\begin{table}[ht!]
\renewcommand{\arraystretch}{1.1}
\centering
\caption{Signal and background samples used in the analysis, together with their fully hadronic production cross sections and expected event yields for an integrated luminosity of 10.8~\invab at \sqrtsZH. The quoted cross sections include the hadronic decay modes of the vector bosons.}
\label{tab:samples}
{\small
\begin{tabular}{lcc}
\toprule
Process & Cross section  & Expected events \\
\midrule
$\rm ZH(\HZZ)$   & 1.73~fb  & $1.87\times10^{4}$ \\
$\rm ZH(\HWW)$   & 13.3~fb  & $1.44\times10^{5}$ \\
$\rm WW$   & 7.47~pb  & $8.07\times10^{7}$ \\
$\rm ZZ$   & 0.665~pb & $7.18\times10^{6}$ \\
$\rm Z/\gamma^*$ & 55.01~pb & $5.94\times10^{8}$ \\
\bottomrule
\end{tabular}
}
\end{table}

\section{Detector response} \label{detresp}

Among the detector concepts proposed for FCC-ee, this analysis uses a modified implementation of the Innovative Detector for Electron-positron Accelerator (IDEA) concept~\cite{IDEA1,IDEA2}. The detector comprises a five-layer silicon pixel vertex detector surrounded by a large-volume, low-mass drift chamber with up to 112 sensitive layers, providing quasi-continuous tracking within a 2~T solenoidal magnetic field. In the IDEA detector configuration considered here, the electromagnetic calorimeter component of the dual-readout system is replaced by a crystal calorimeter to provide improved electromagnetic energy resolution, while the hadronic calorimeter retains the dual-readout design. The detector is completed by an outer system of $\mu$-RWELL chambers for muon identification.

Detector simulation and reconstruction are performed with \DELPHES~\cite{deFavereau:2013fsa}, using a parameterized description of the detector response to emulate the expected resolutions. The tracking system covers $|\cos\theta|<0.988$, while the electromagnetic and hadronic calorimeters extend to $|\cos\theta|<0.995$. The output consists of particle-flow candidates, which are used as input to the jet clustering and subsequent event reconstruction. The full chain of event generation, detector simulation, and reconstruction is integrated within the \KEYfHEP software framework~\cite{key4hep}.

The expected performance of the hadronic dual-readout calorimeter is modeled in \DELPHES through the energy-resolution parameterization
\begin{equation}
  \rm \left(\frac{\sigma_E}{E}\right)^2
  =
  (1\%)^2
  + \frac{(30\%)^2}{E/\GeV}
  + \frac{(5\%)^2}{(E/\GeV)^2},
  \label{eq:calo_resolution}
\end{equation}
where the three contributions correspond to the constant, stochastic, and noise terms, respectively. The parameterization is applied in both the central and forward calorimeter regions, yielding a relative hadron energy resolution of approximately $4.4\%$ at 50~GeV. For jets in the energy range relevant to this analysis, the particle-flow jet momentum resolution is found to be close to that of the neutral-hadron component, which provides the dominant detector contribution through the hadronic calorimeter response.

\section{Preselection and event reconstruction}\label{preselection}

The preselection is designed to select events with fully hadronic activity while suppressing backgrounds containing genuine missing momentum or isolated leptons. Events containing an identified electron or muon with $\rm p>10$~GeV are vetoed, reducing semileptonic and leptonic backgrounds and ensuring orthogonality with other \HWW or \HZZ analyses focusing on final states containing charged leptons. The reconstructed missing momentum, defined as the negative vector sum of the momenta of all reconstructed particles in the event, is required to satisfy $\rm p_{\rm miss}<20$~GeV.

\begin{figure}[t!]
  \centering
    \includegraphics[width=0.95\linewidth]{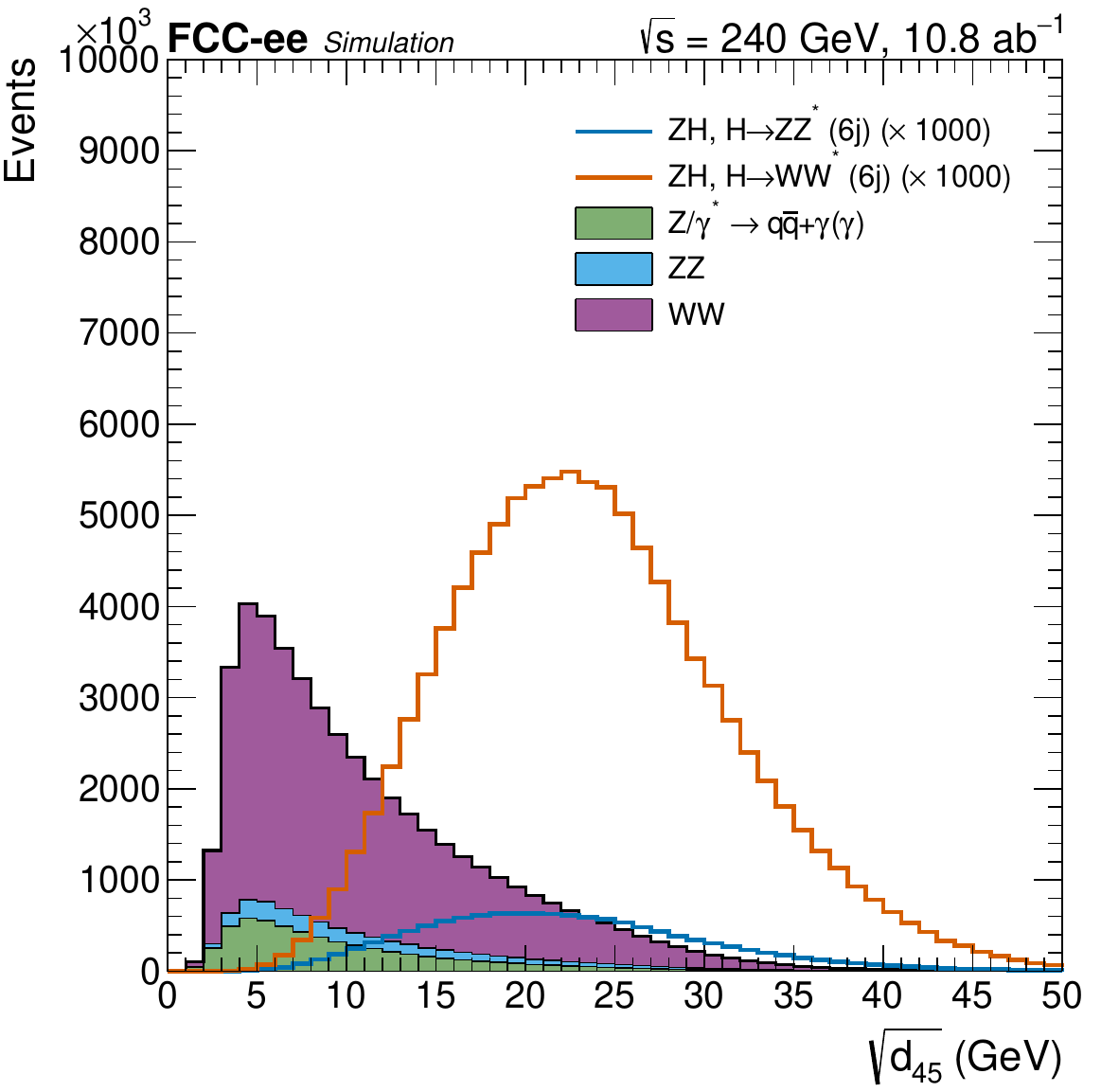}
  \caption{Distribution of the jet-transition scale $\rm \sqrt{d_{45}}$ for signal and background processes. For visualization purposes, the signal distributions are scaled by a factor of 1000.}
\label{fig:dmerge45}
\end{figure}

The remaining events are clustered exclusively into six jets using the Durham $\rm k_T$ algorithm~\cite{Catani:1991hj} as implemented in \FASTJET~\cite{fastjet}. Events for which six valid jets cannot be reconstructed are discarded. In addition to the reconstructed jet four-momenta, the Durham clustering sequence provides the jet-transition scales $\rm d_{ij}$, which characterize the resolution scale at which an event changes from a $j$-jet to an $i$-jet configuration (e.g. $\rm d_{56}$ corresponds to the transition from six to five jets). At tree level, the signal processes contain six quarks in the final state, whereas the dominant diboson backgrounds contain four primary quarks, and the $\rm e^+e^-\to Z/\gamma^*\to q\bar q$ background contains only two. Additional jets reconstructed in these lower-multiplicity backgrounds typically arise from softer or more collinear QCD radiation and are therefore associated with smaller jet-transition scales. Fig.~\ref{fig:dmerge45} illustrates the resulting separation in the jet-transition scale $\rm \sqrt{d_{45}}$, which characterizes the transition between five- and four-jet configurations. The six-parton signal processes populate larger values of $\rm \sqrt{d_{45}}$ than the lower-multiplicity backgrounds. To exploit these differences, the requirements $\sqrt{d_{23}}>50$~GeV, $\sqrt{d_{34}}>25$~GeV, $\sqrt{d_{45}}>15$~GeV, and $\sqrt{d_{56}}>10$~GeV are imposed, with the threshold values determined from an optimization of the expected signal significance.

\begin{figure}[ht!]
  \centering
    \includegraphics[width=0.95\linewidth]{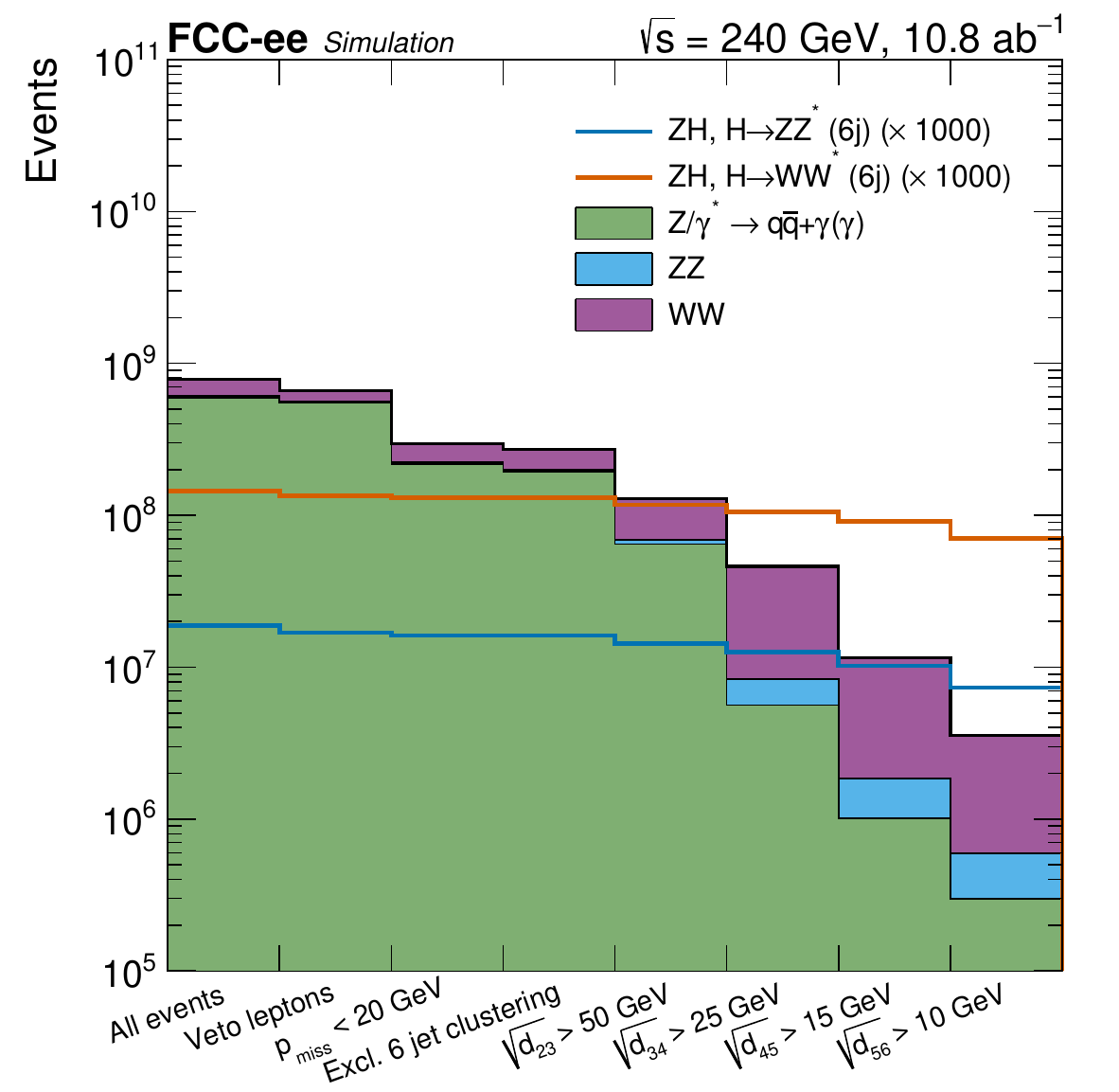}
  \caption{Cutflow for signal and background processes after the successive preselection requirements. The ``All events'' bin corresponds to the initial fully hadronic signal samples, inclusive WW and ZZ background samples, and hadronic fermion-pair production, before any selection requirements. For visualization purposes, the signal yields are scaled by a factor of 1000.}
\label{fig:cutflow}
\end{figure}

The cumulative effect of the preselection is illustrated by the cutflow in Fig.~\ref{fig:cutflow}, with the corresponding event yields and cumulative selection efficiencies summarized in Tables~\ref{tab:cutflow} and~\ref{tab:cutflow_eff}, respectively. Relative to the initial event samples, approximately 39\% of the \HZZ signal and 49\% of the \HWW signal remain after the full preselection, while about 98\% of both the $\rm WW$ and $\rm ZZ$ inclusive backgrounds are rejected. The rejection is particularly strong for the fermion-pair background, for which more than 99.9\% of the events are removed. The jet-transition requirements provide a substantial fraction of this discrimination, with approximately 46\% of the \HZZ and 54\% of the \HWW events passing these selections, while, relative to the six-jet requirement, about 96\% of the $\rm WW$ background, 95\% of the $\rm ZZ$ background, and more than 99.8\% of the fermion-pair background are rejected. The lower preselection efficiency for the \HZZ signal primarily results from the joint optimization of the \HWW and \HZZ selections, while the softer kinematics of the off-shell $\rm Z^*$ system further contribute to the more challenging reconstruction of \HZZ events. The selection cuts, particularly those based on the jet-transition variables, are chosen to maximize the combined \HWW+\HZZ signal significance. Since the expected \HWW yield is substantially larger than that of \HZZ, the resulting working point is naturally more favorable to the \HWW topology, leading to a modestly lower \HZZ efficiency.

\begin{table}[ht]
\renewcommand{\arraystretch}{1.1}
\centering
\caption{Event yields after the successive selection requirements. The initial yields correspond to the fully hadronic signal samples, inclusive WW and ZZ background samples, and hadronic fermion-pair production. The final rows correspond to the event categorization described in Section~\ref{hzz_hww_sep}.}
\label{tab:cutflow}
\footnotesize
\setlength{\tabcolsep}{3.5pt}
\begin{tabular}{@{}lccccc@{}}
\hline
Selection &
\HZZ &
\HWW &
WW &
ZZ &
$\rm Z/\gamma^{*}$ \\
&
(6j) &
(6j)  &
(incl.) &
(incl.) &
(2q) \\
&
\multicolumn{2}{c}{$(\times 10^{3})$} &
\multicolumn{3}{c}{$(\times 10^{6})$} \\
\hline
All events
    & 18.7 & 144  & 178  & 14.7  & 594 \\
Lepton veto
    & 16.8 & 135  & 100  & 10.5  & 553 \\
$\rm p_{\mathrm{miss}}$
    & 16.1 & 130  & 72.0 & 5.89  & 217 \\
6 jets
    & 16.1 & 130  & 71.7 & 5.40  & 194 \\
$\sqrt{d_{23}}$
    & 14.4 & 117  & 58.9 & 4.38  & 64.6 \\
$\sqrt{d_{34}}$
    & 12.5 & 105  & 37.7 & 2.72  & 5.63 \\
$\sqrt{d_{45}}$
    & 10.2 & 91.6 & 9.70 & 0.838 & 1.01 \\
$\sqrt{d_{56}}$
    & 7.35 & 70.5 & 2.96 & 0.295 & 0.298 \\
\hline
\HZZ cat.
    & 3.96 & 10.5 & 0.433 & 0.112 & 0.08 \\
\HWW cat.
    & 3.39 & 60.0 & 2.52 & 0.183 & 0.215 \\
\hline
\end{tabular}
\end{table}

\begin{table}[ht]
\renewcommand{\arraystretch}{1.1}
\centering
\caption{Cumulative selection efficiencies (in \%) relative to the initial event samples. The signal samples correspond to the fully hadronic \HWW and \HZZ final states, while the WW and ZZ backgrounds are inclusive and the fermion-pair background is hadronic. The final rows correspond to the event categorization described in Section~\ref{hzz_hww_sep}.}
\label{tab:cutflow_eff}
\footnotesize
\setlength{\tabcolsep}{4pt}
\begin{tabular}{@{}lccccc@{}}
\hline
Selection &
\HZZ &
\HWW &
WW &
ZZ &
$\rm Z/\gamma^{*}$ \\
\hline
All events
    & 100.0 & 100.0 & 100.0 & 100.0 & 100.0 \\
Lepton veto
    & 89.8 & 93.8 & 56.2 & 71.4 & 93.1 \\
$\rm p_{\mathrm{miss}}$
    & 86.1 & 90.3 & 40.4 & 40.1 & 36.5 \\
6 jets
    & 86.1 & 90.3 & 40.3 & 36.7 & 32.7 \\
$\sqrt{d_{23}}$
    & 77.0 & 81.3 & 33.1 & 29.8 & 10.9 \\
$\sqrt{d_{34}}$
    & 66.8 & 72.9 & 21.2 & 18.5 & 0.95 \\
$\sqrt{d_{45}}$
    & 54.5 & 63.6 & 5.45 & 5.70 & 0.17 \\
$\sqrt{d_{56}}$
    & 39.3 & 49.0 & 1.66 & 2.01 & 0.05 \\
\hline
\HZZ cat.
    & 21.2 & 7.29 & 0.24 & 0.76 & 0.013 \\
\HWW cat.
    & 18.1 & 41.7 & 1.42 & 1.24 & 0.036 \\
\hline
\end{tabular}
\end{table}

\section{Separation of \HWW and \HZZ} \label{hzz_hww_sep}

The six jets in each event are grouped into three pairs, corresponding to the on-shell vector boson from the Higgs decay, the off-shell vector boson, and the Z boson from the associated ZH production. The optimal jet assignment is determined by minimizing the following $\chi^2$-like quantity
\begin{equation} 
\rm \chi^2 = (m_{ij}-m_V)^2 + (m_{kl}-m_Z)^2 + (m_{ijmn}-m_H)^2 \,, \label{eq:jetpairing} 
\end{equation} 

where $\rm m_{ij}$ is the invariant mass of the jet pair assigned to the on-shell vector boson from the Higgs decay, $\rm m_{kl}$ is that of the jet pair assigned to the associated Z boson, and $\rm m_{ijmn}$ is the invariant mass of the four jets assigned to the Higgs boson candidate. The remaining jet pair, $\rm (mn)$, is identified with the off-shell vector boson and its invariant mass is therefore left unconstrained. The indices $\rm i,j,k,l,m,n$ run over the six reconstructed jets, with each jet entering exactly one pair. All 90 possible jet-pairing assignments are considered, and the assignment minimizing the $\chi^2$ is selected. The parameter $\rm m_V$ is set to the nominal W or Z boson mass for the \HWW and \HZZ hypotheses, respectively, while $\rm m_Z$ and $\rm m_H$ denote the nominal Z and Higgs boson masses. Since the relevant reconstructed mass resolutions are dominated by the hadronic energy resolution and are assumed to be comparable, a common resolution factor is omitted from Eq.~\eqref{eq:jetpairing}. For each selected event, the minimization is performed independently under the \HWW and \HZZ hypotheses, resulting in two corresponding jet assignments and sets of reconstructed mass observables. 

\begin{figure}[t!]
  \centering
  \includegraphics[width=0.49\linewidth]{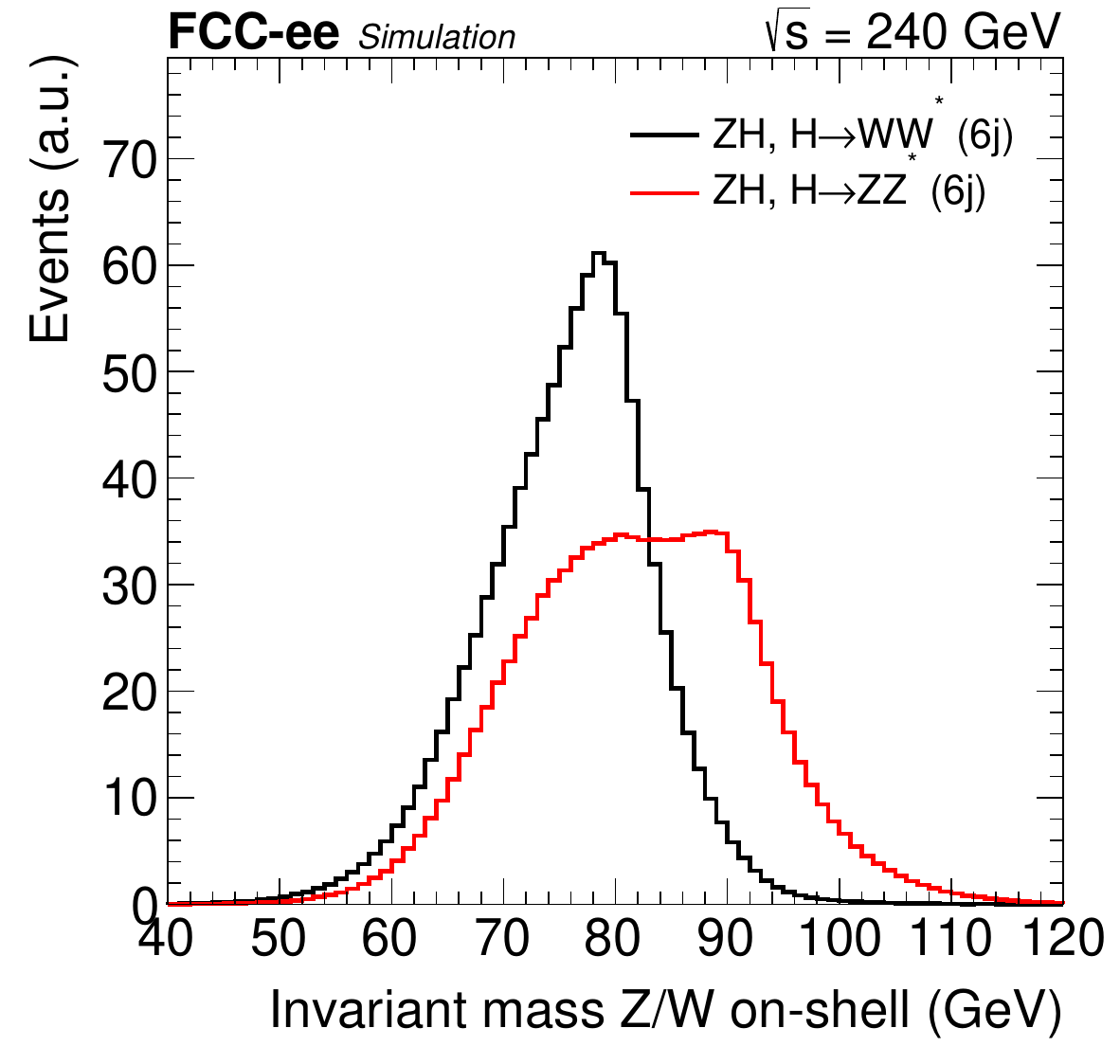}
  \includegraphics[width=0.49\linewidth]{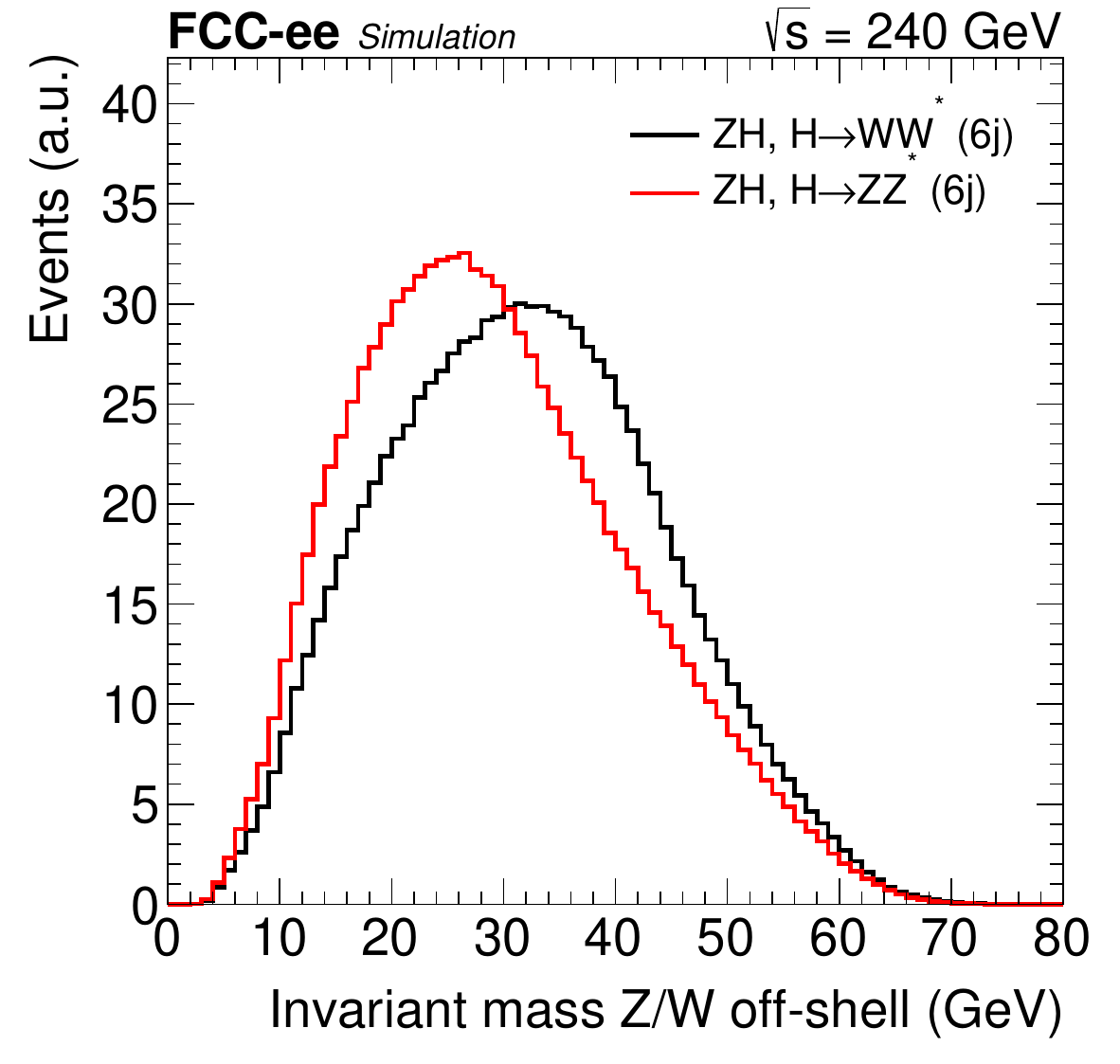}
  \includegraphics[width=0.49\linewidth]{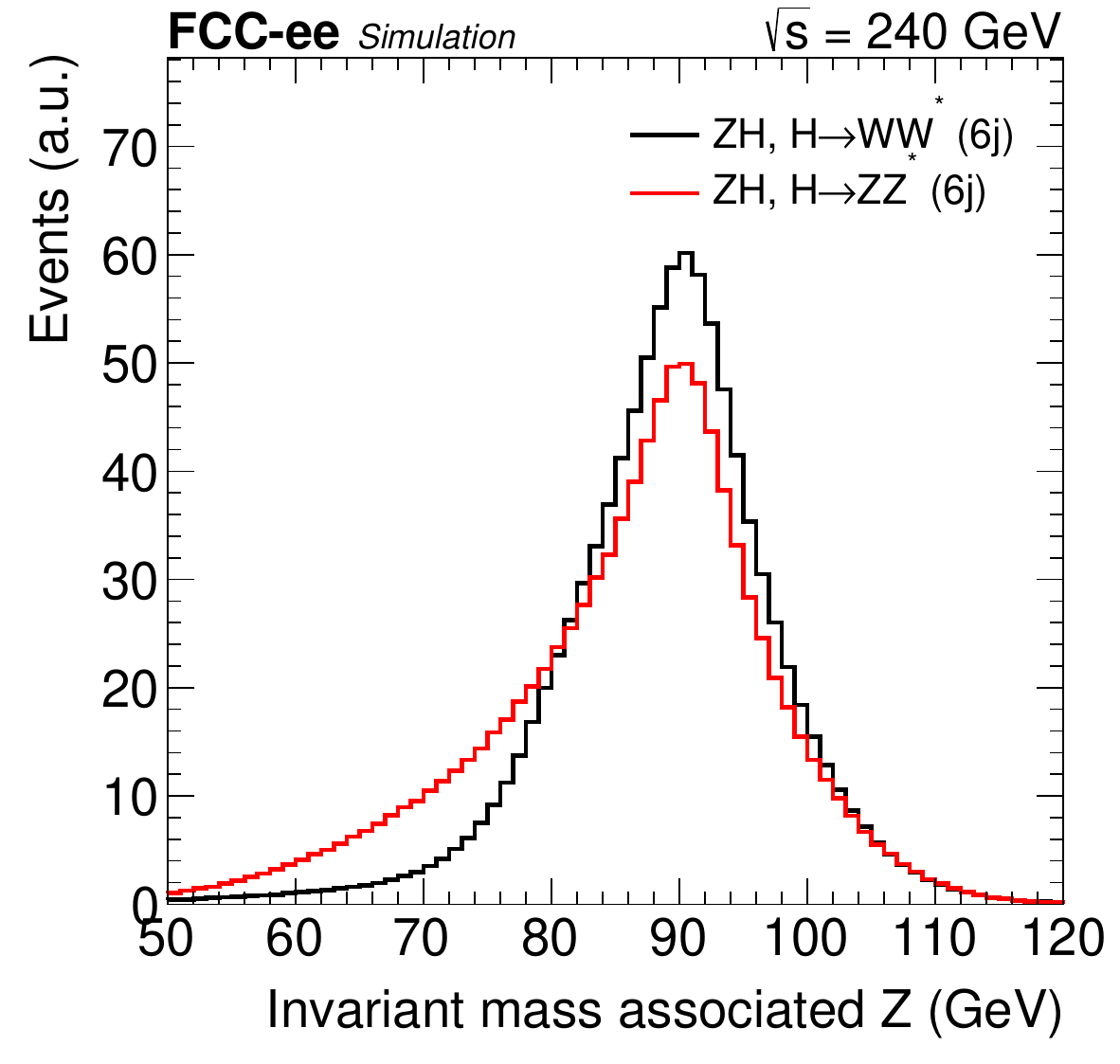}
  \includegraphics[width=0.49\linewidth]{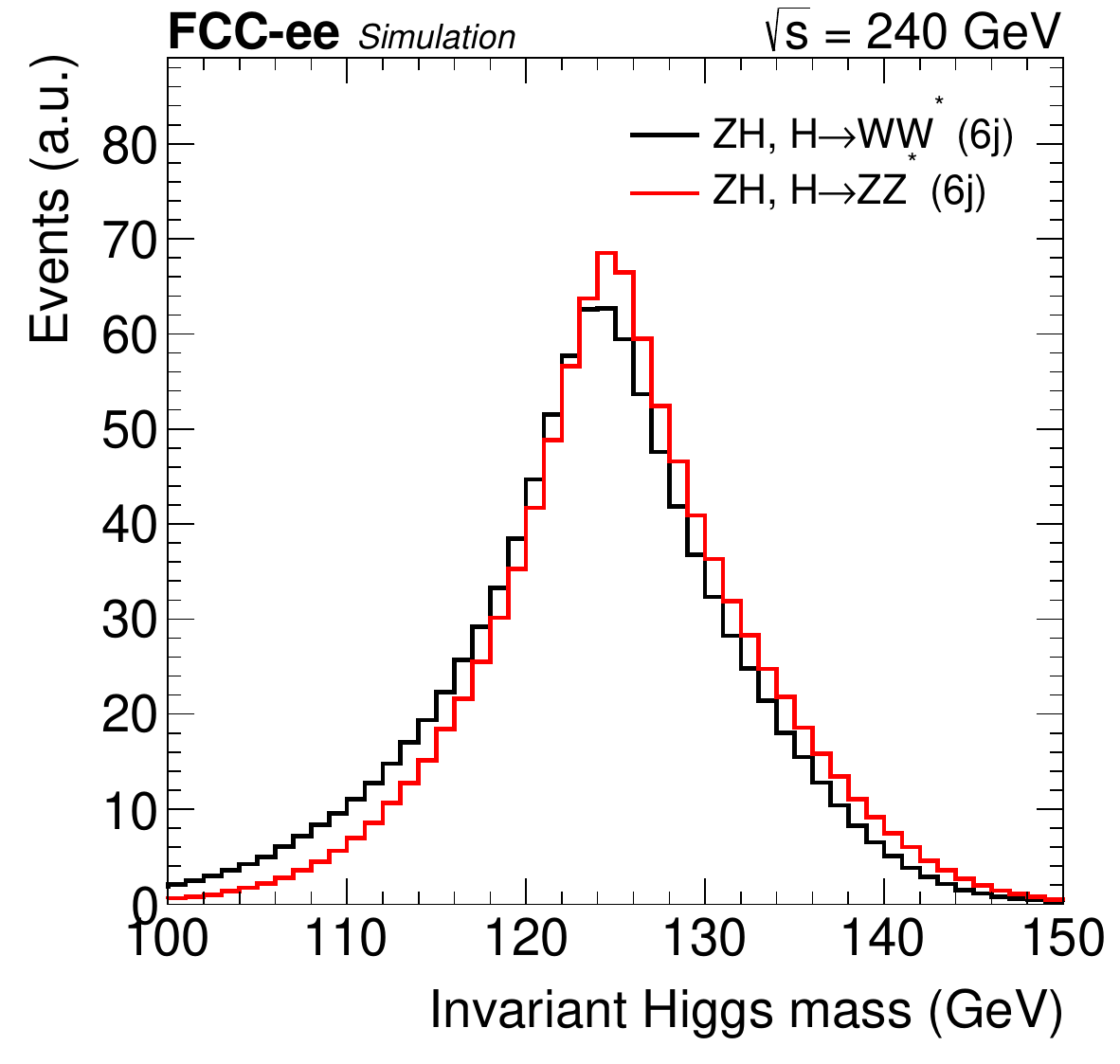}
  \caption{Invariant mass distributions resulting from the jet-pairing $\rm \chi^2$ minimization for \HWW (black) and \HZZ (red) signal events: on-shell vector boson (top left), off-shell vector boson (top right), associated Z boson (bottom left), and reconstructed Higgs boson (bottom right).}
\label{fig:jet_pairing}
\end{figure}

The performance of this procedure is illustrated in Fig.~\ref{fig:jet_pairing}, which shows the reconstructed Higgs, associated-Z, on-shell-V, and off-shell-V masses for the \HWW and \HZZ signal samples. For \HWW events, the procedure provides a good reconstruction of the Higgs boson, the associated Z boson, and the on-shell W boson, while the reconstructed $\rm W^{*}$ mass exhibits the broad distribution expected for an off-shell decay. The reconstruction performance is less favorable for \HZZ events, as the Higgs boson and associated Z boson masses are well reconstructed, but the on-shell Z boson originating from the Higgs decay is systematically reconstructed below the nominal Z-boson mass. This behavior indicates a larger ambiguity in the partition of the four Higgs-decay jets between the Z and $\rm Z^{*}$ candidates. The lower characteristic invariant mass of the off-shell $\rm Z^{*}$ system compared with the $\rm W^{*}$ system further increases the sensitivity of this assignment to jet-clustering and combinatorial effects. Further optimization of the jet pairing, for example through the direct inclusion of jet-flavor or multivariate information in the jet pairing procedure, could improve the pairing efficiency, particularly for the \HZZ hypothesis. In the present analysis, jet-flavor information is instead exploited at a later stage (see below) in the multivariate classifier used to separate \HWW and \HZZ events. A constrained kinematic fit exploiting the known Z- and Higgs-boson masses could additionally improve the hadronic mass resolution and the discrimination between the two hypotheses. Such refinements of the jet-pairing procedure are not considered here.

\begin{figure}[t!]
  \centering
  \includegraphics[width=0.49\linewidth]{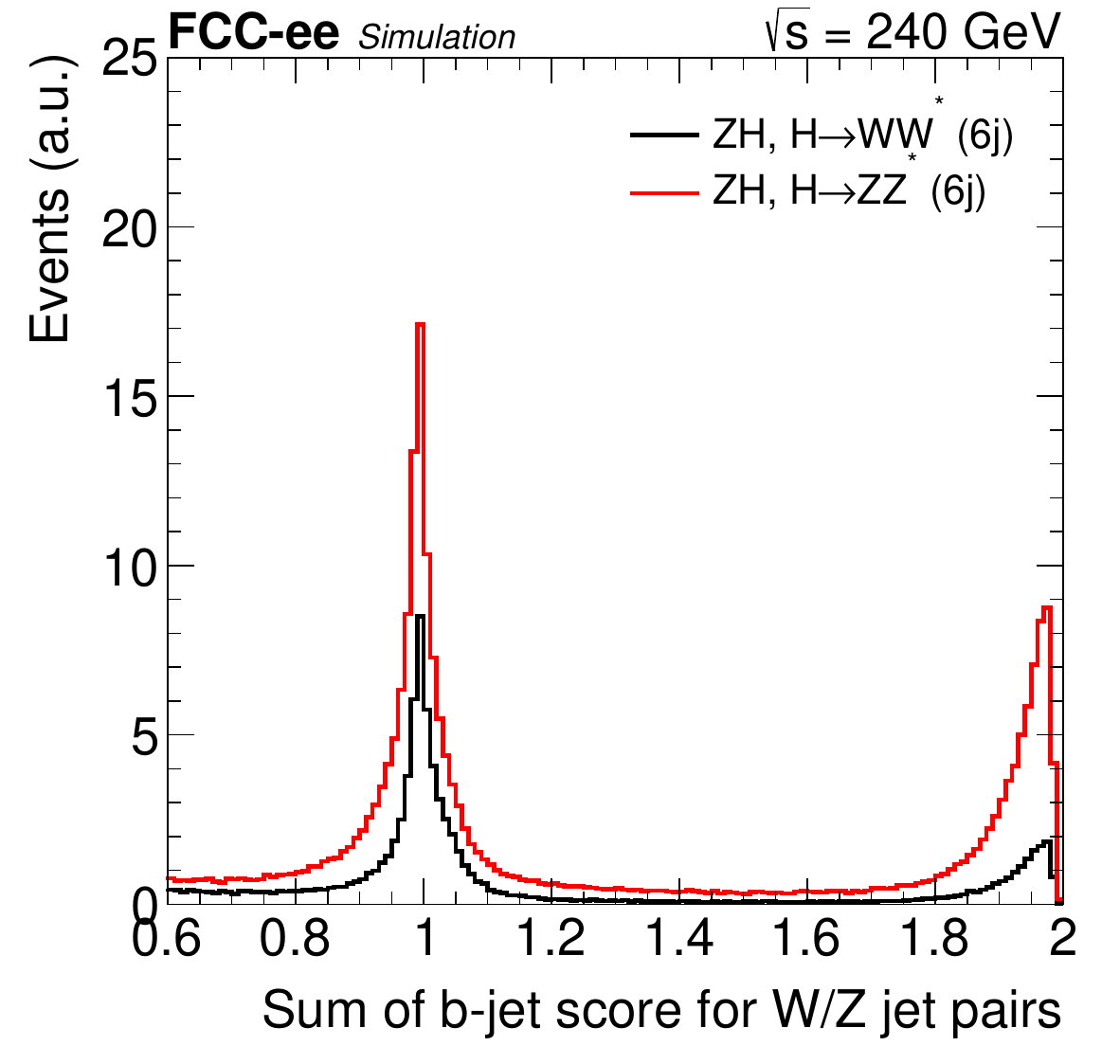}
  \includegraphics[width=0.49\linewidth]{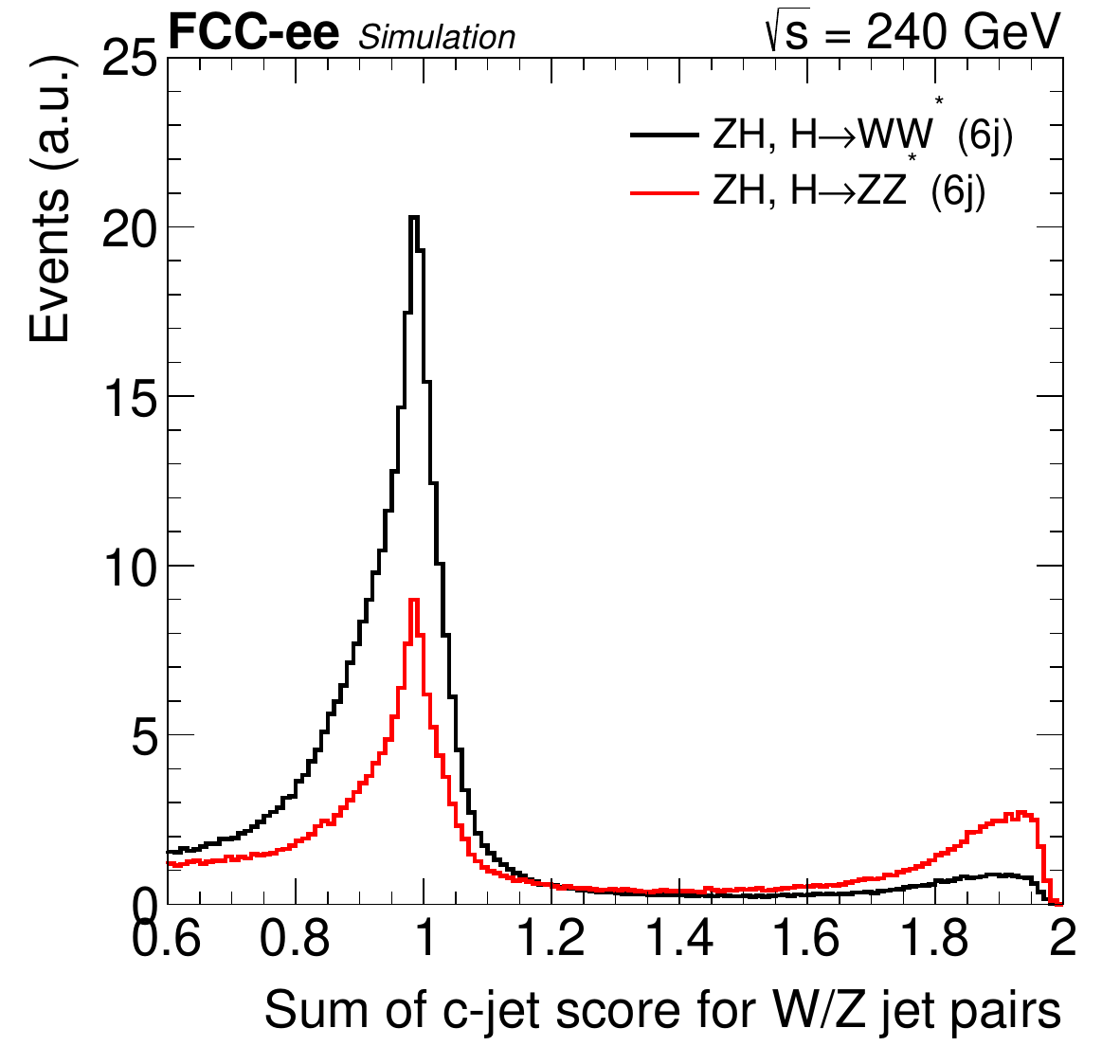}
  \caption{Sum of the $b$-jet (left) and $c$-jet (right) flavor-tagging scores of the two jets assigned to the on-shell vector-boson candidate for \HWW and \HZZ events. The different heavy-flavor composition of hadronic W- and Z-boson decays provides discrimination between the two signal processes.}
\label{fig:jet_pairing_flavor}
\end{figure}

To discriminate between the \HWW and \HZZ decay modes, a Boosted Decision Tree (BDT) classifier is trained using \HWW events as signal and \HZZ events as background. The input variables include jet kinematic observables, including the jet momenta and polar angles, as well as the reconstructed invariant masses of the vector-boson and Higgs-boson candidates obtained under both the \HWW and \HZZ jet-pairing hypotheses described above (see Fig.~\ref{fig:jet_pairing}). Jet-flavor information provides additional discrimination because hadronic W-boson decays necessarily involve quarks of different flavor, whereas Z-boson decays produce same-flavor quark pairs. This information is incorporated using the Graph Neural Network (GNN) jet-flavor tagger~\cite{Bedeschi:2022rnj}. For each jet, the tagger provides probabilities associated with seven flavor hypotheses: gluon (g), up (u), down (d), strange (s), charm (c), bottom (b), and tau ($\tau$). The tagger exploits information from the particle-flow candidates associated with each jet, including their kinematic properties, impact-parameter information, and particle-identification observables such as cluster counting and time-of-flight measurements. The flavor information provides some of the leading discriminating variables of the BDT. Fig.~\ref{fig:jet_pairing_flavor} shows, as examples, the sums of the b- and c-jet scores of the two jets assigned to the on-shell vector-boson candidate. The \HZZ process exhibits a more pronounced peak near a summed flavor score of two due to $\rm Z\to b\bar{b}$ and $\rm Z\to c\bar{c}$ decays, in which both jets originate from heavy-flavor quarks. In contrast, hadronic W-boson decays can contain at most one heavy-flavor quark, except for the strongly CKM-suppressed $\rm W \rightarrow c\bar{b}$ decay. These differences in flavor composition provide complementary information to the reconstructed boson masses and kinematic observables for separating the two Higgs decay modes.

\begin{figure}[t!]
  \centering
    \includegraphics[width=0.95\linewidth]{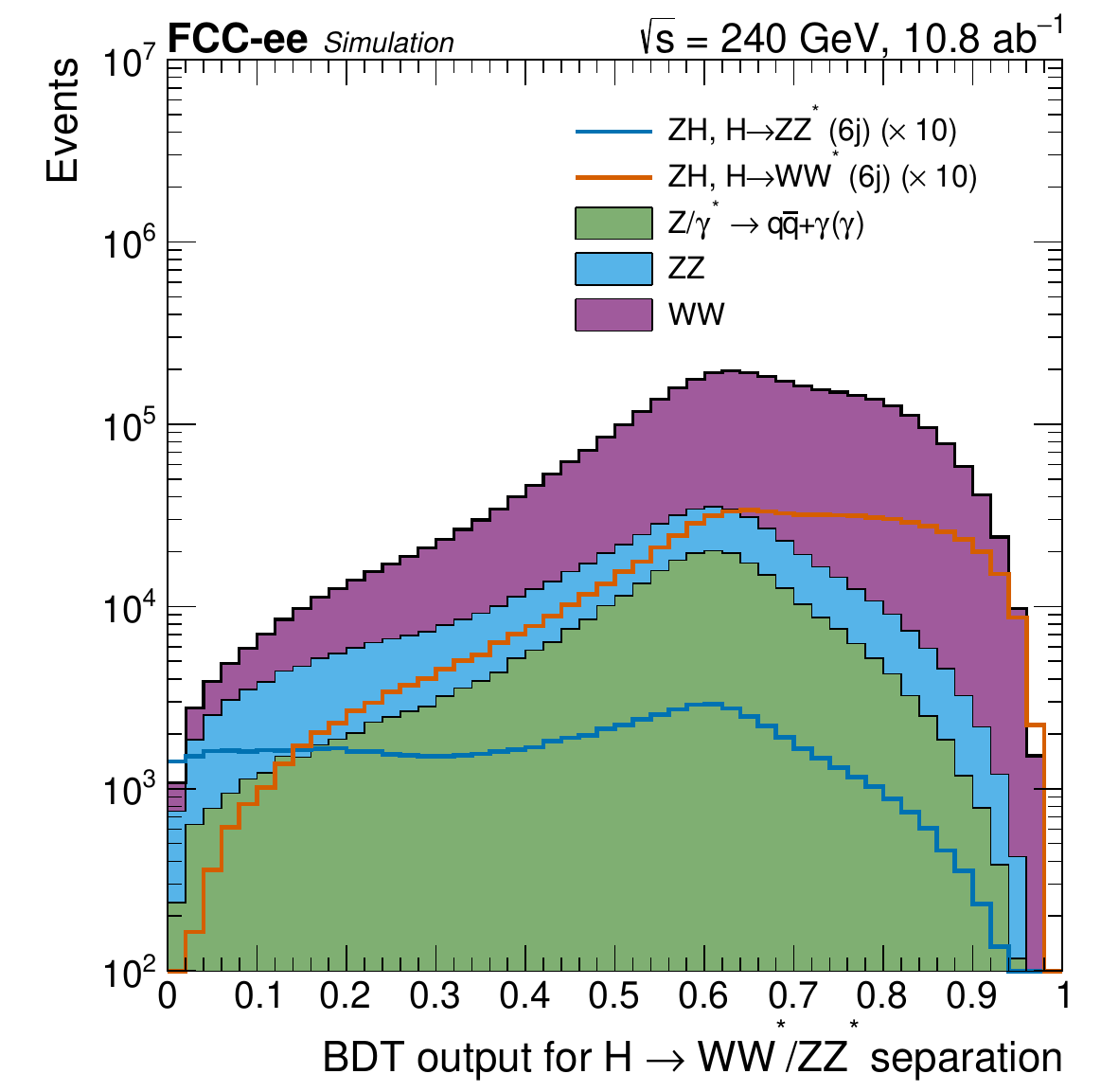}
  \caption{BDT output distributions for the \HWW and \HZZ signal processes and the backgrounds. The classifier is trained using \HWW as signal and \HZZ as background. Events with a BDT score above (below) $0.5$ are classified as \HWW-like (\HZZ-like). The signal distributions are scaled by a factor of 10 for visibility.}
\label{fig:mva_output_hww_hzz}
\end{figure}

After training, the BDT achieves an area under the ROC curve of 0.77, with the reconstructed vector-boson masses and the jet-flavor scores associated with the \HWW pairing hypothesis providing the largest discrimination. The resulting BDT output distributions for the signal and background processes are shown in Fig.~\ref{fig:mva_output_hww_hzz}, with the corresponding event yields in the two categories summarized in Table~\ref{tab:cutflow}. Events with a BDT score below $0.5$ are classified as \HZZ-like, while those with a score above $0.5$ are classified as \HWW-like. The same categorization is applied to all background events, resulting in two mutually exclusive event categories. With this working point, approximately $85\%$ of \HWW events are classified as \HWW-like, while the corresponding classification efficiency for \HZZ events is approximately $54\%$. The lower classification efficiency for the \HZZ mode is consistent with the larger combinatorial ambiguities observed in the jet pairing and the poorer reconstruction of the on-shell Z boson.

As illustrated in Fig.~\ref{fig:mva_output_hww_hzz}, the background processes preferentially populate the \HWW-like category, with approximately 85\%, 62\%, and 73\% of the selected WW, ZZ, and \Zg events assigned to this category, respectively. This behavior is consistent with the kinematic and flavor-tagging information used by the BDT, which is trained exclusively to discriminate between the two Higgs decay modes. In particular, the WW background shares the characteristic hadronic W-boson decay kinematics and flavor composition of the \HWW signal, while jet-clustering and pairing effects can lead to similar reconstructed observables for the other background processes.

\section{Multivariate background suppression}\label{optimize}

After the categorization into \HWW-like and \HZZ-like events, the two categories are optimized independently to suppress the remaining backgrounds. A dedicated BDT is trained in each category, using the corresponding Higgs decay mode as signal and all remaining background processes as background. The dominant background contribution in both categories arises from $\rm WW$ production.

The same set of input observables is used for the classifiers in both categories, while the jet-pairing hypothesis is chosen consistently with the targeted signal process. The variables include the momenta and polar angles of the reconstructed jets, the kinematic properties of the reconstructed boson candidates, including the invariant masses and momenta of the Higgs boson and the on- and off-shell vector bosons, as well as the jet-flavor information introduced in Section~\ref{hzz_hww_sep}. These observables provide additional discrimination between the signal and the remaining backgrounds after the event categorization.

\begin{figure}[ht!]
  \centering
  \includegraphics[width=0.95\linewidth]{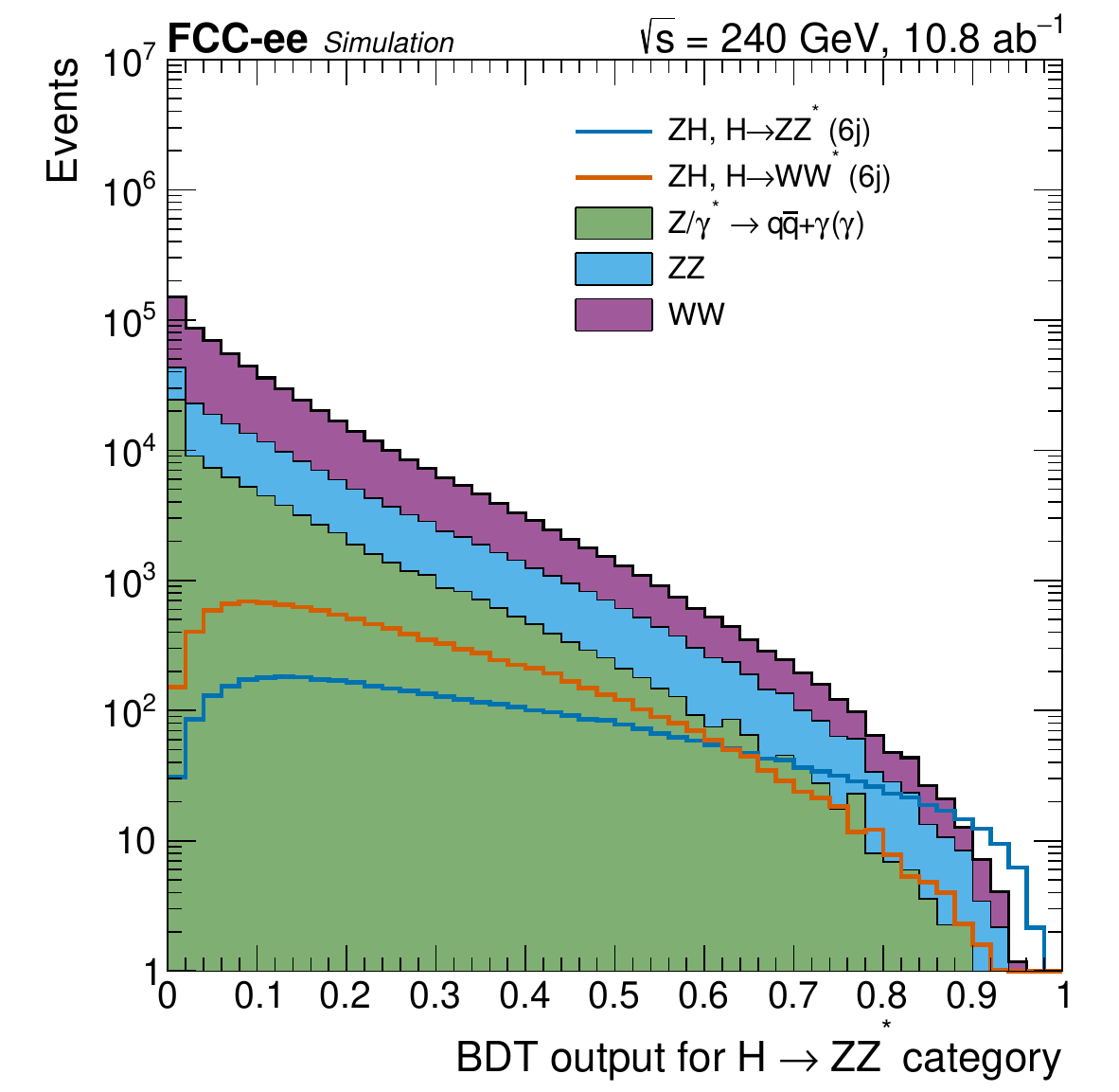}
  \includegraphics[width=0.95\linewidth]{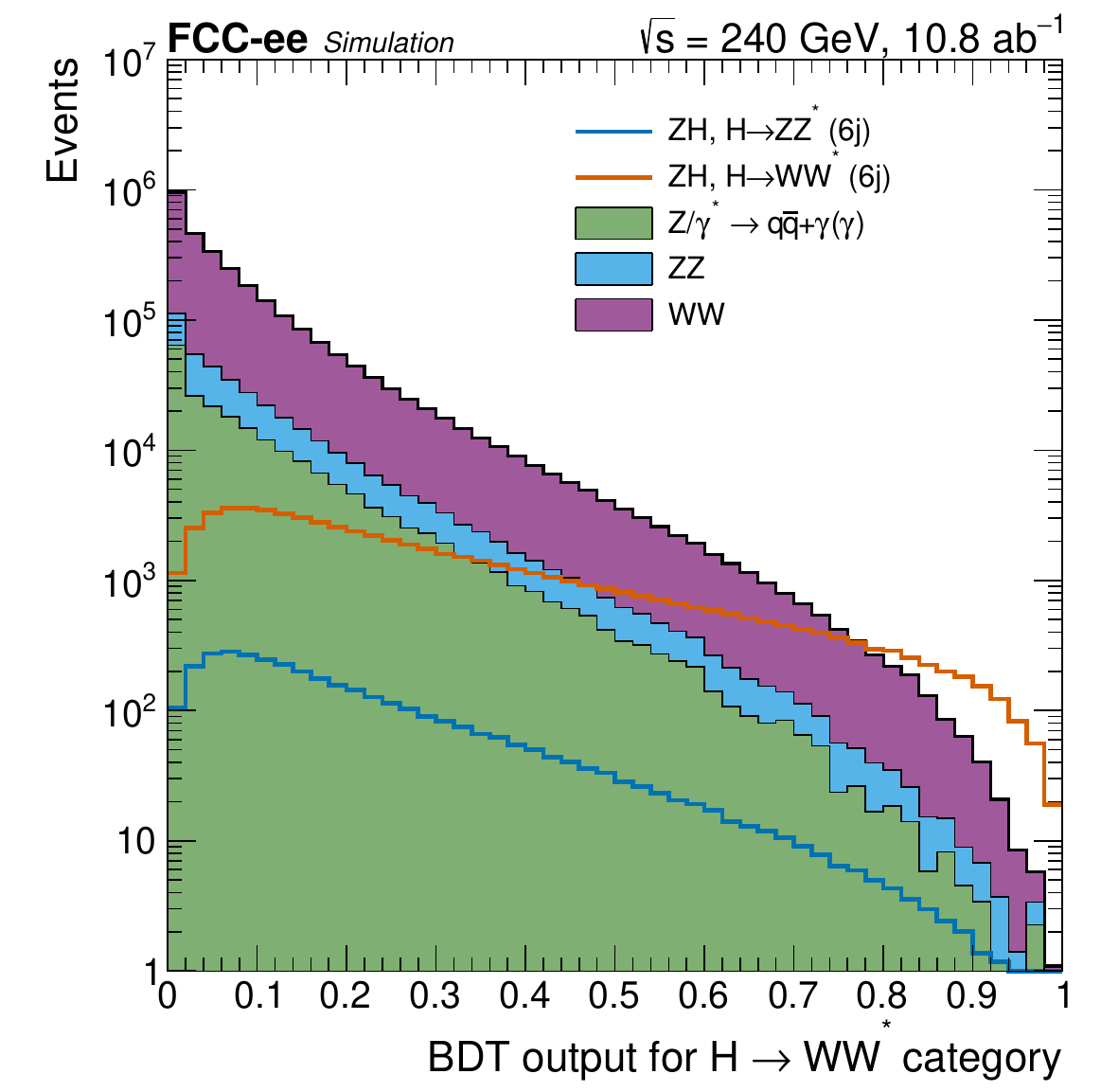}
  \caption{Output distributions of the dedicated BDT classifiers used for background suppression in the \HZZ-like (top) and \HWW-like (bottom) categories. In both categories, the background contribution decreases towards larger classifier scores, resulting in an enhanced signal-to-background ratio at high values.}
\label{fig:mva_hww_hzz}
\end{figure}

The resulting BDT output distributions for the \HZZ-like and \HWW-like categories are shown in Fig.~\ref{fig:mva_hww_hzz}. In both categories, the background contribution is progressively suppressed towards larger BDT scores, resulting in an enhanced signal-to-background ratio at high classifier values. Rather than applying an additional selection on the classifier output, the full distributions are retained and used as inputs to the simultaneous binned likelihood fit described in Section~\ref{results}.

\section{Results} \label{results}

The output discriminants of the \HWW and \HZZ analyses (see Fig.~\ref{fig:mva_hww_hzz}) are used as inputs to a simultaneous binned likelihood fit to extract the corresponding signal strengths, defined as
\begin{equation}
\rm
\mu_X =
\frac{\sigma_{\mathrm{ZH}}\times\mathrm{BR}(H\to X)}
     {[\sigma_{\mathrm{ZH}}\times\mathrm{BR}(H\to X)]_{\mathrm{SM}}},
\qquad X = WW^*, ZZ^*.
\end{equation}
Both signal strengths are treated as independent parameters of interest and are allowed to float freely in the fit. Expected sensitivities are evaluated using an Asimov dataset constructed from the SM signal and background predictions. A 1\% normalization uncertainty is assigned to each background process, correlated between the two event categories. This is taken as a conservative benchmark, given the expected theoretical precision and the large control samples available to constrain the background rates directly from data.

\begin{figure}[t!]
  \centering
    \includegraphics[width=0.95\linewidth]{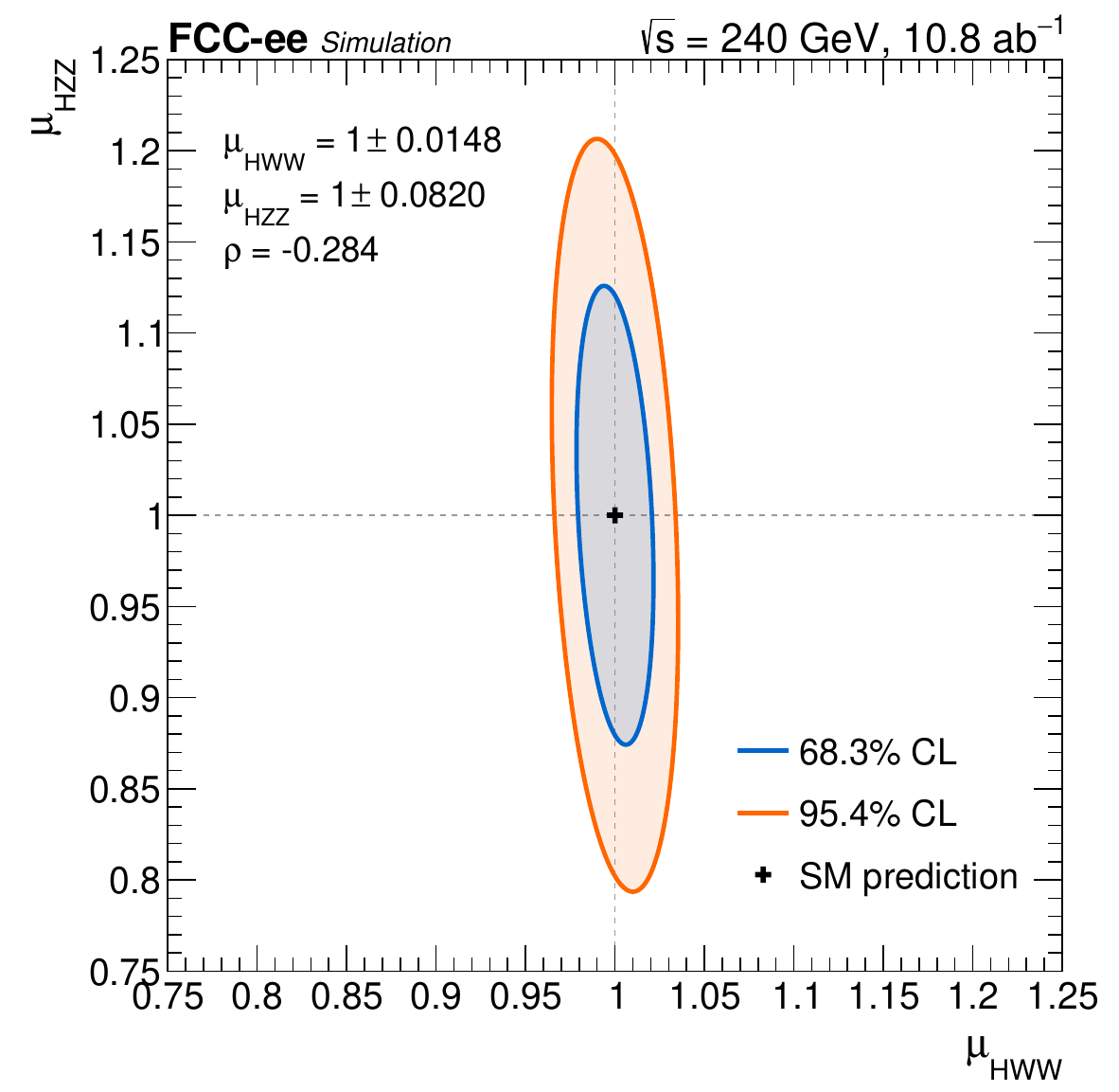}
  \caption{Expected 68.3\% and 95.4\% confidence contours for the simultaneous measurement of the $\rm H\to WW^*$ and $\rm H\to ZZ^*$ signal strengths. The contours are derived from the covariance matrix of the maximum-likelihood fit.}
\label{fig:contour}
\end{figure}

The simultaneous fit yields expected relative uncertainties of 1.48\% and 8.20\% on the \HWW and \HZZ signal strengths, respectively. The two parameters exhibit a correlation of $-28.4\%$, as shown in Fig.~\ref{fig:contour}, reflecting the residual cross-contamination between the \HWW and \HZZ categories. When the background-normalization uncertainties are removed, the corresponding expected uncertainties improve to 1.32\% and 7.78\%, respectively. As a further test of the impact of the mutual \HWW--\HZZ contamination, fixing the \HZZ contribution when extracting the \HWW signal strength improves the expected uncertainty from 1.48\% to approximately 1.42\%, while fixing the \HWW contribution in the \HZZ fit improves the corresponding uncertainty from 8.20\% to approximately 7.92\%.

These results can be compared with the combined FCC-ee projections of approximately 0.8\% for $\rm \sigma_{ZH}\times BR(\HWW)$ and 2.5\% for $\rm \sigma_{ZH}\times BR(\HZZ)$, when combining the significant final states studied so far~\cite{Selvaggi:2025kmd}. Complementary \HWW studies with leptonic or invisible decays of the associated Z boson and fully hadronic Higgs decays reach individual precisions of approximately 1.6\% and 1.3\%, respectively. For \HZZ, a wider range of leptonic, semileptonic, and invisible final states has been investigated, with individual sensitivities ranging from a few percent to the $\mathcal{O}(10\%)$ level. When the individual channels are combined independently, adding the fully hadronic channels studied in this work improves the combined precision by approximately 16\% for \HWW and 5\% for \HZZ. The fully hadronic final states considered here therefore provide a significant contribution to the overall FCC-ee sensitivity to both decay modes.

Apart from the background-normalization uncertainties described above, systematic uncertainties related to the detector response and signal modeling are not explicitly included in the present analysis. For the fully hadronic six-jet final states, the most relevant experimental effects are expected to arise from the jet energy scale and resolution, jet clustering and particle assignment, and jet-flavor tagging performance. In addition to the jet energy resolution, combinatorial effects arising from the assignment of the six jets to the vector bosons also contribute to the broadening of the reconstructed mass distributions. Imperfect modeling of these effects may alter the reconstructed boson-mass and multivariate-discriminant distributions, affecting the separation of the \HWW and \HZZ decay modes. However, these effects can be constrained in situ using the abundant $\rm Z/\gamma^*\to q\bar q$, $\rm WW$, and $\rm ZZ$ control samples available at FCC-ee. The integrated luminosity is expected to be known at the per-mille level at FCC-ee from Bhabha and di-photon events, while theoretical uncertainties become relevant in the interpretation of the measured rates in terms of Higgs couplings. A complete assessment of detector, modeling, and theoretical uncertainties will require dedicated studies. Nevertheless, the availability of abundant control samples for constraining the dominant experimental effects, together with the modest degradation observed when including background-normalization uncertainties, suggests that the dominant systematic uncertainties are expected to remain below the statistical uncertainty and that the analysis therefore has the potential to remain predominantly statistically limited.

\section{Conclusion}  \label{conclusion}

We have studied the sensitivity to the \HWW and \HZZ decay modes in associated Higgs-boson production at FCC-ee, focusing on the fully hadronic final state in which the associated Z boson and the vector bosons from the Higgs decay all decay to quarks. The analysis assumes an integrated luminosity of 10.8~\invab at \sqrtsZH, corresponding to the baseline FCC-ee program with four interaction points, and an IDEA-like detector performance. The fully hadronic topology benefits from the large branching fractions of the hadronic W- and Z-boson decays, but presents significant challenges due to large multijet backgrounds, combinatorial ambiguities in the reconstruction of the boson candidates, and the substantial overlap between the \HWW and \HZZ final states.

A dedicated analysis strategy is developed in which events are first selected using their characteristic six-jet topology and subsequently separated into \HWW-like and \HZZ-like categories using a multivariate classifier combining kinematic and jet-flavor information. Dedicated multivariate discriminants are then used to suppress the remaining backgrounds in each category. A simultaneous binned likelihood fit to the two discriminant distributions yields expected relative uncertainties of 1.48\% on $\rm \sigma_{ZH}\times BR (\HWW)$ and 8.20\% on $\rm \sigma_{ZH}\times BR (\HZZ)$, including a 1\% normalization uncertainty on each background process.

Detector and signal modeling systematic uncertainties are not explicitly evaluated in this study. The dominant experimental effects, related to the jet energy response, jet reconstruction, and flavor tagging, can be constrained using the abundant hadronic control samples available at FCC-ee. Together with the modest impact of the background-normalization uncertainties, this indicates that the analysis is expected to remain predominantly statistically limited.

These results demonstrate that the fully hadronic final state provides significant sensitivity to both the \HWW and \HZZ decay modes and can make an important contribution to the precision determination of the Higgs couplings to electroweak gauge bosons and, ultimately, to the model-independent determination of the Higgs boson width at FCC-ee.

\begin{acknowledgements}
The work of C. Paus and J. Eysermans is supported by the U.S. Department of Energy, Office of Science, Office of High Energy Physics under contract no. DE‐SC0011939.
\end{acknowledgements}

\bibliographystyle{spphys}
\bibliography{references}

\end{document}